\documentclass[12pt]{article}
\usepackage{amssymb,amsmath}
\usepackage[colorlinks=true, linkcolor=blue, citecolor=blue, urlcolor=blue]{hyperref}
\def\a{\alpha}
\def\b{\beta}

\def\g{\gamma}

\def\be{\begin{equation}}
\def\ee{\end{equation}}
\def\arr{\begin{array}{rll}}
\def\ea{\end{array}}
\def\bea{\begin{eqnarray}}
\def\eea{\end{eqnarray}}
\newcommand{\nn}{\nonumber}

\def\N2{$N{=}2$}

\def\>{\rangle}
\def\<{\langle}
\def\+{\dagger}
\def\={\ =\ }

\def\pa{{\partial}}

\def\cal#1{\mathcal{#1}}

\title{Eisenhart Lift of $2$--Dimensional Mechanics}
\author{Allan P. Fordy\thanks{School of Mathematics,
University of Leeds, Leeds LS2 9JT, UK. ~~E-mail: a.p.fordy@leeds.ac.uk}
$\,$ and Anton Galajinsky\thanks{School of Physics, Tomsk Polytechnic University, 634050 Tomsk, Russia. ~~E-mail: galajin@tpu.ru}
}

\begin{document}

\maketitle
\bibliographystyle{plain}

\begin{abstract} \noindent
The Eisenhart lift is a variant of geometrization of classical mechanics with $d$ degrees of freedom in which the equations of motion are embedded into the geodesic equations of a Brinkmann–-type metric defined on $(d+2)$--dimensional spacetime of Lorentzian signature.
In this work, the Eisenhart lift of $2$--dimensional mechanics on curved background is studied. The corresponding $4$--dimensional metric is governed by two scalar functions which are just the conformal factor and the potential of the original dynamical system. We derive a conformal symmetry and a corresponding quadratic integral, associated with the Eisenhart lift. The energy--momentum tensor is constructed which, along with the metric, provides a solution to the Einstein equations. Uplifts of $2$--dimensional superintegrable models are discussed with a particular emphasis on the issue of hidden symmetries. It is shown that for the $2$--dimensional Darboux--Koenigs metrics, only type I can result in Eisenhart lifts which satisfy the weak energy condition. However, some physically viable metrics with hidden symmetries are presented.
\end{abstract}


\medskip



\section{Introduction}

It is known since Eisenhart's work on the geometrization of classical mechanics \cite{LE} that any dynamical system with $d$ degrees of freedom $q^i$, $i=1,\dots,d$, which is governed by the Lagrangian $\mathcal{L}$, can be embedded into the geodesic equations of the Brinkmann–-type metric $2 \mathcal{L} dt^2-dt dv$, where $t$ is the temporal variable and $v$ is an extra coordinate. When analyzing the geodesic equations, one finds that
$t$ is affinely related to the proper time $\tau$, the equations of motion for $q^i(t)$ coincide with those following from the Lagrangian $\mathcal{L}$, while the evolution of $v(t)$ is unambiguously fixed provided $q^i(t)$ are known. The initial dynamical system is thus recovered by implementing the null reduction along $v$.

Curiously enough, the original publication of \cite{LE} did not receive much attention by physicists and had
soon fallen into oblivion. Several decades passed before the method was rediscovered in \cite{DBKP}, which paved the way for various physical applications (see \cite{DGH} and references therein).

Particularly interesting geometries result from uplifts of integrable and superintegrable systems. Constants of the motion polynomial in momenta give rise to Killing vectors and Killing tensors, the rank of the latter being equal to the degree of the polynomial.  Killing vectors are associated with a clear statement of symmetry. They result from infinitesimal coordinate transformations which leave the form of a metric invariant. On the other hand, higher order Killing tensors (associated with {\em hidden symmetries}) have no such simple interpretation.  {\em Second order} Killing tensors are associated with separation of variables of the Hamilton-Jacobi equation, with Carter's integration of the geodesic equations in the Kerr metric \cite{68-4} being the prime example in General Relativity.  Although quite a few physically meaningful spacetimes have been constructed, which admit one or several second rank Killing tensors (for a recent review see \cite{FKK}), no solution to the vacuum Einstein equations admitting higher rank Killing tensors is presently known.\footnote{Ricci--flat metrics of ultrahyperbolic signature, which admit higher rank Killing tensors, were constructed in \cite{CG}.} This empirical barrier of rank--2 seems rather puzzling. It may be a technical issue but perhaps something more fundamental lies behind it. The study of Lorentzian metrics admitting higher rank Killing tensors within Eisenhart's approach generated an extensive recent literature \cite{GHKW}--\cite{CGHHZ}. While the geometric reformulation of Newtonian mechanics brings mostly aesthetic advantages, the construction of Brinkmann--type metrics with hidden symmetries is a source of new results.

Thus far attention was mostly drawn to integrable mechanics in flat space (for some curved space examples see \cite{GHKW,GR}). Although the interrelationship between geometric characteristics of the Eisenhart metric and those of a Riemannian metric underlying mechanics on a curved background is generally rather complicated, the analysis is greatly simplified for $2D$ case, because $2$--dimensional manifolds are conformally flat.

In recent years there has been a burst of activity in the identification and classification of superintegrable systems, both classical and quantum (see the review \cite{MPW} and references therein).
Most of the interest is in Hamiltonians which are in ``natural form'' (the sum of kinetic and potential energies), with the kinetic energy being {\em quadratic} in momenta and therefore associated with a (pseudo-)Riemannian metric.  When an $n$--dimensional space is either flat or constant curvature, it possesses the maximal group of isometries, which is of dimension $\frac{1}{2}n(n+1)$.  In this case, the kinetic energy is actually the second order {\em Casimir} function of the symmetry algebra (see \cite{74-7}).  Furthermore {\em all} higher order integrals of the geodesic equations are built out of the corresponding Noether constants by just taking polynomial expressions in them.  Whilst most of the classification results and examples which occur in applications correspond to flat or constant curvature spaces, there are well known examples of conformally flat spaces (but {\em not} constant curvature), possessing quadratic invariants, which are clearly not just quadratic expressions in Noether constants.  Specifically, there are the metrics found by Koenigs \cite{K}, which are described and analysed in \cite{02-6,KKMW}.  There are other examples of conformally flat spaces (but {\em not} constant curvature), possessing one Noether constant and a {\em cubic} integral (classified in \cite{11-3} and further studied and generalised in \cite{VDS,V}), which again {\em cannot} be represented as a cubic expression in the isometry algebra. Being conformally flat, these spaces {\em do have} an abundant supply of {\em conformal symmetries} and in \cite{AF,FH} a method was proposed for building quadratic and higher order {\em invariants} from appropriate polynomial expressions in {\em conformal invariants}.

The goal of this paper is to study the Eisenhart lift of $2$--dimensional mechanics in curved space. Specifically, we consider the relationship of the curvature of the $2$--, and $4$--dimensional geometries and the structure of the Einstein tensor.  We also consider the conformal symmetries of the Eisenhart lift and use this to build an additional quadratic invariant.  We are particularly interested in constructing physically admissible energy--momentum tensors in a purely geometric way and to derive equations which connect geometric characteristics to those of matter.

The work is organized as follows.

In Section \ref{eis-lift}, the equations of motion of a generic $2$--dimensional mechanical system in curved space are embedded into the geodesic equations of a Brinkmann type $4$--dimensional metric. The latter is determined by two scalar functions which are just the conformal factor and the potential of the original $2D$ mechanics.  We see that first integrals can always be lifted from the $2$ to the $4$--dimensional domain, with the addition of 2 further involutive integrals, thus preserving Liouville integrability. In Section \ref{conform-sym} we show that the Eisenhart lift has a conformal symmetry for a large class of $2D$ Hamiltonians.  This conformal symmetry is then used to construct a new first integral for the Eisenhart lift.

In Section \ref{Energy-Momentum} we discuss the energy--momentum tensor of the $4$--dimensional lift. Restrictions on the energy--momentum tensor, which follow from the weak and strong energy conditions, are formulated. In Section \ref{hidden} we discuss some specific Hamiltonian systems, comparing Liouville integrability with superintegrability. It is shown that $2D$ Darboux--Koenigs metrics result in $4D$ solutions which violate the weak energy condition in types II, III, and IV, but obey it in the invariant region $x>0$ for type I. A physically viable metric, admitting a rank 2 Killing tensor, is constructed by uplifting a superintegrable model on $S^2$, as well as other models with the additional functional freedom of being just Liouville integrable.

Some final remarks are gathered in the concluding Section \ref{conclude}.

\section{Eisenhart Lift of $2$--Dimensional Mechanics in Curved Space}\label{eis-lift}

In this section we describe the Eisenhart lift, which is similar to the Kaluza--Klein extension, which allows us to consider the motion of a particle in a curved background, under the influence of a potential, as {\em geodesic} motion on a larger curved space.

First we give a brief review of ideas from geometric mechanics and the relationship between first integrals and Killing tensors.

\subsection{First Integrals in Classical Mechanics}\label{first-int}

A particle moving in a curved background, with metric coefficients $g_{ij}$ and scalar potential $U({\bf q})$, can be written in Lagrange form, with ${\cal L} = \frac{1}{2} \sum_{i,j=1}^n g_{ij}\dot q^i \dot q^j -U({\bf q})$ or, equivalently, in Hamiltonian form with $H = \frac{1}{2} \sum_{i,j=1}^n g^{ij} p_i p_j+U({\bf q})$, where $g^{ij}$ are {\em inverse} metric coefficients and $p_i=\sum_{j=1}^n g_{ij}\dot q^j$. In either case, we obtain
\be\label{geq}
\frac{d^2q^k}{dt^2} + \sum^n_{i,j=1}\Gamma^k_{ij}\frac{dq^i}{dt}\frac{dq^j}{dt} = -\sum^n_{i=1}g^{ki}\pa_i U   \, , \quad k=1,\dots,n ,
\ee
where $\Gamma^k_{ij}$ are the components of the Levi-Civita connection.  The simplest way to compute the connection coefficients is to use the Hamiltonian to calculate $\{\{q^k,H\},H\}$ and to replace $p_i$ by the above formula and then to just equate coefficients to those in (\ref{geq}).

In the Lagrangian framework we have Noether's theorem, which relates symmetries of the Lagrangian to first integrals (Noether constants).  This connection is more transparent in the Hamiltonian formulation, being related to first integrals which are of first degree in momenta.  In the case of geodesic equations, there is a direct relation to Killing vectors.  Given
\begin{subequations}
\be\label{HandK}
H_0 = \frac{1}{2} \sum_{i,j=1}^n g^{ij} p_ip_j \quad\mbox{and}\quad  K = \sum_{i=1}^n a^i({\bf q}) p_i,
\ee
then
\be\label{KH0}
\{K,H_0\} = 0 \quad\Rightarrow\quad  \nabla_{(i}a_{j)}=0,
\ee
where $\nabla_i$ denotes the covariant derivative.  Furthermore, Hamilton's equations for $K$ generate a {\em vector field on configuration space}:
\be\label{Kv}
\hat K = \sum_{i=1}^n a^i({\bf q}) \frac{\pa}{\pa q^i},
\ee
\end{subequations}
since the first $n$ components are written entirely in terms of the {\em position} variables.  $\hat K$ is just the Killing vector corresponding to the Noether constant $K$.

Similarly, if $F=\sum_{i,j=1}^n f^{ij}p_ip_j$, $f^{ij}$ the components of a symmetric matrix, then
\be\label{HandF}
\{F,H_0\} = 0 \quad\Rightarrow\quad  \nabla_{(i}f_{jk)}=0,
\ee
thus defining a rank 2 Killing tensor.  In general, if $F$ is an integral which is homogeneously polynomial of degree $m$ in momenta, then the coefficients define a Killing tensor of rank $m$.

Such integrals no longer generate a vector field, such as (\ref{Kv}), on configuration space, so have a different geometric significance.  A Killing \underline{vector} defines a ``motion'' on configuration space, defined by the dynamical system
$$
\frac{d q^i}{ds} = a^i({\bf q}),
$$
under which the metric is invariant.  For simple cases, such as rotations, this can be explicitly solved to give a transformation of coordinates.  For each Killing vector $K$ it is possible ({\em in principle}) to find new coordinates such that $K=\frac{\pa}{\pa Q_1}$, in which case the metric coefficients are {\em independent} of the variable $Q_1$.  If we have $n$ {\em commuting} Killing vectors $K_i$ (the flat case), then it is possible to find coordinates $Q_1,\dots ,Q_n$, such that $K_i=\frac{\pa}{\pa Q_i}$, in which case metric coefficients are constant.

Whilst {\em quadratic} integrals have no such simple geometric meaning, they arise in the theory of separation of variables of the Hamilton-Jacobi equation.  In fact, a {\em complete solution} of the Hamilton-Jacobi equation depends upon $n$ parameters, which can then be written in terms of the dynamical variables to give $n$ mutually commuting {\em quadratic} integrals (some of which could be squares of {\em linear} integrals).

Whilst $n$ is the maximal number of independent functions which can be {\em in involution}, it is possible to have further integrals of the Hamiltonian $H$, which necessarily generate a non-Abelian algebra of integrals of $H$.  The maximum number of additional {\em independent} integrals is $n-1$, since the "level surface" of $2n-1$ integrals (meaning the intersection of individual level surfaces) is just the (unparameterised) integral curve.  Such systems are called {\em superintegrable} ({\em maximal} when there are $2n-1$ independent integrals).  Well known elementary examples are the isotropic harmonic oscillator, the Kepler system and the Calogero-Moser system.  The role of superintegrability in both the classical and quantum context is described in \cite{f18-3}.  It is much stronger than just Liouville integrability.  In this paper we discuss both Liouville and superintegrable examples, particularly in the context of satisfying the weak energy condition.

\subsection{Conformally Flat Spaces in $2$ Dimensions}\label{conform-flat}

The Lagrangian kinetic energy is directly related to the metric $ds^2 = \sum_{i,j=1}^n g_{ij}dq^idq^j$.  In 2 dimensions, all metrics are conformally flat, so there exist coordinates $(q^1,q^2)=(x,y)$, with respect to which the metric takes the form
\begin{subequations}
\be\label{2dMetric}
ds^2 = \frac{dx^2+dy^2}{\varphi(x,y)} ,
\ee
with only one independent curvature component, given by the {\em scalar curvature}
\be\label{2dR}
R = \varphi \Delta \log(\varphi),
\ee
where $\Delta f = f_{xx}+f_{yy}$.  The Eisenhart lift of this metric incorporates the potential function $U(x,y)$ in the definition of ${\cal L}$:
\be\label{4dMetric}
ds^2 = \frac{dx^2+dy^2}{\varphi(x,y)}-dv dt - 2 U(x,y) dt^2.
\ee
We require both $\varphi(x,y)$ and $U(x,y)$ to be strictly positive, at least in some region of the $(x,y)$ plane which is invariant under the geodesic flow.
We also require that the signature of the metric is $(+,+,+,-)$.

Since the $2D$ metric satisfies the Einstein vacuum equations, the Einstein tensor for the Eisenhart lift has {\em at most} only 3 non-zero components, which are easily computed:
\be\label{4dEinstein}
R_{ij}-\frac{1}{2} R g_{ij} = \left(
                                \begin{array}{cccc}
                                  0 & 0 & 0 & 0 \\
                                  0 & 0 & 0 & 0 \\
                                  0 & 0 & 0 & \frac{1}{4} R \\
                                  0 & 0 &  \frac{1}{4} R & \varphi \Delta U+R U
                                \end{array}
                              \right),
\ee
with $R$ taking the save value (\ref{2dR}).  These can be equated with the energy momentum tensor, which naturally lies in the tensor spaces $\pa_t\otimes\pa_v+\pa_v\otimes\pa_t$ and $\pa_t\otimes\pa_t$.
\end{subequations}

\subsection{The Hamiltonian Formulation of the Eisenhart Lift}\label{Hamilton-2d4d}

The Hamiltonian in 2 dimensions takes the form
\begin{subequations}
\be\label{GenH2d}
H^{2d} = \frac{1}{2} \varphi(x,y) (p_x^2+p_y^2) +U(x,y).
\ee
The Eisenhart lift of this Hamiltonian is
\be\label{GenH}
H = \frac{1}{2} \varphi(x,y) (p_x^2+p_y^2) +4 U(x,y) p_v^2- 2 p_vp_t.
\ee
\end{subequations}
In the $2$--dimensional context the symbol "$t$" represents the Hamiltonian "time-parameter", with $\frac{d x}{dt} = \frac{\pa H^{2d}}{\pa p_x}$, etc, whilst for the Eisenhart lift, $t$ is a {\em coordinate} in spacetime, whose evolution in the new "time parameter" $\tau$ can be found by using Hamilton's equations from (\ref{GenH}):
\begin{subequations}\label{dbydtau}
\bea
&& \frac{dx}{d\tau} =  \varphi p_x,\quad \frac{dy}{d\tau} = \varphi p_y,\quad
     \frac{dv}{d\tau} = 8 U  p_v-2 p_t,\quad  \frac{dt}{d\tau} =  -2 p_v,  \label{dxdtau}  \\
&&  \frac{dp_x}{d\tau} = -\frac{1}{2} \varphi_x (p_x^2+p_y^2)-4 U_x p_v^2,\quad   \frac{dp_y}{d\tau} = -\frac{1}{2} \varphi_y (p_x^2+p_y^2)-4 U_y p_v^2,\label{dpxdtau}  \\
  &&    \frac{dp_v}{d\tau} = 0,\quad   \frac{dp_t}{d\tau} =  0.   \label{dtdtau}
\eea
\end{subequations}
Equations (\ref{dtdtau}) immediately give that $p_v$ and $p_t$ are first integrals, so we can set
\begin{subequations}
\bea
  p_v=-\frac{1}{2} \kappa,  &\quad\Rightarrow\quad &  \frac{dt}{d\tau} = \kappa,\label{pv}  \\
  p_t = - \kappa {\cal E} &\quad\Rightarrow\quad &   \frac{dv}{dt} + 4 U = 2 {\cal E},\label{pt}
\eea
where $\kappa$ and ${\cal E}$ are constants.
The evolution of $(x,y,p_x,p_y)$ just gives the original equations associated with the $2$--dimensional Hamiltonian (\ref{GenH2d}).

If, in 2 dimensions, there is a quadratic first integral of (\ref{GenH2d}):
\be\label{F2d}
F^{2d} = \sum_{i,j=1}^2 f^{ij}(x,y)p_ip_j+w(x,y), \quad\mbox{with}\quad \{H^{2d},F^{2d}\}=0,
\ee
then, in 4 dimensions, this is "homogenised" to give
\be\label{F4d}
F = \sum_{i,j=1}^2 f^{ij}(x,y)p_ip_j+4 w(x,y)p_v^2, \quad\mbox{satisfying}\quad \{H,F\}=0,
\ee
\end{subequations}
and $\{p_v,F\}=\{p_t,F\}=0$, so $H, F, p_v,p_t$ are in involution.  Thus Liouville integrability is preserved in this construction (see also \cite{GHKW}).

In a similar way we can extend higher order integrals from 2 dimensions to the $4$--dimensional Eisenhart lift.

\section{A Conformal Invariant in $4D$ with an Additional Quadratic Invariant}\label{conform-sym}

Whilst the $2$--dimensional metric is conformally flat (with an infinite number of conformal Killing vectors), the $4$--dimensional extension (\ref{4dMetric}) is not (unless the $2D$ curvature $R=0$ and $U(x,y)$ takes a very simple form).  For the Hamiltonian (\ref{GenH}) we ask that there exists a first degree (in momenta) {\em conformal invariant}
\be\label{GenK}
K = a_1 p_x+a_2 p_y+a_3 p_v+a_4 p_t, \quad\mbox{satisfying}\quad \{K,H\} = \sigma(x,y,v,t) H,
\ee
where $a_1, a_2, a_3$ are functions of $(x,y,v,t)$, whilst $a_4(t)$ is a function of only $t$.
This equation is homogeneously quadratic in momenta, so gives us 10 equations for the coefficients $a_i$, which can be partially solved.  We quickly find that $a_1$ and $a_2$ can be written in terms of a potential $w(x,y,v,t)$, with $a_1=w_x, a_2= -w_y$ and with $w$ satisfying the $2$--dimensional Laplace equation
\begin{subequations}\label{w-eq}
\be\label{laplace}
w_{xx}+w_{yy}=0.
\ee
The equations are generally fairly complicated, but simplified by assuming that $w$ is independent of $v$ and $t$.

First the coefficients of $p_xp_v$ and $p_yp_v$ give us formulae for $w_{xxx}$ and $w_{xxy}$, which are compatible ($(w_{xx})_{xy}=(w_{xx})_{yx}$ is {\em identically} satisfied) and can be integrated to give
\be\label{wxx}
w_{xx} = \frac{1}{2} (w_x \pa_x-w_y \pa_y)\log(\varphi)+b =  \frac{1}{2} \{\log(\varphi),K\}+b,
\ee
after which we have explicit formulae for $a_3$ and $a_4$, giving\footnote{Here and below $b$ and $c_i$ designate constants.}
\be\label{Ksol}
K = w_x p_x-w_y p_y +((2 b-c_4)v+c_3 t)p_v + c_4 t p_t,
\ee
with
\be\label{KH}
\{K,H\}=2 b H,
\ee
and $U(x,y)$ satisfying
\be\label{Ueq}
\{U,K\}=(w_x \pa_x-w_y \pa_y)U = 2 (b-c_4)U -\frac{1}{2} c_3.
\ee
\end{subequations}

Since we have $\{t,H\}=-2 p_v$ and (\ref{KH}), it follows that
\begin{subequations}
\be\label{FK}
F_K = K p_v+ b\, t\, H
\ee
is a first integral.  This is in addition to $p_v, p_t$ and any first integrals inherited from the $2$--dimensional system.  These additional first integrals satisfy
\be\label{FKpbs}
\{p_v,p_t\} = 0, \quad  \{p_v,F_K\} = (c_4-2 b)p_v^2,\quad \{p_t,F_K\} = -(c_3 p_v+c_4 p_t)p_v-b H.
\ee
\end{subequations}

\subsection{Some Solutions of the System (\ref{w-eq})}\label{Examples}

We can read Equations (\ref{w-eq}) in two ways.

We can choose a solution of Laplace's equation (\ref{laplace}) to insert into the remaining equations, to be solved for $\varphi(x,y)$ and $U(x,y)$, and then analyse the resulting system.  This analysis could be of the resulting Hamiltonian system and/or the {\em geometry} of the resulting $4$--dimensional metric.  Is it an Einstein vacuum metric?  If not, what is its energy-momentum tensor?

Alternatively, we could start with a $2$--dimensional metric (ie the function $\varphi(x,y)$) and then solve Equations (\ref{w-eq}) for $w(x,y)$, giving the conformal symmetry $K$ (and the value of $b$), and then solve (\ref{Ueq}) for the compatible family of potential functions $U(x,y)$.

\subsubsection{Starting with $w(x,y) = x^2-y^2$} \label{given-w}

With this choice of $w(x,y)$, we easily find that
$$
\varphi(x,y) = x^{2-b} A\left(\frac{y}{x}\right),\quad U(x,y) = \frac{c_3}{4(b-c_4)} +x^{b-c\-4} B\left(\frac{y}{x}\right),
$$
and $K$ takes the form $K=2(x p_x+y p_y)+(c_3t+(2 b-c_4)v)p_v+c_4 t p_t$.  Thus, with this particular $w(x,y)$, we have two arbitrary functions of a single variable in the definition of the metric (\ref{4dMetric}), each with this conformal symmetry $K$.  We look at a particular example in Section \ref{hid-Liouv}.

\subsubsection{Starting with $\varphi(x,y) = 1$} \label{HH-pot}
With this choice of $\varphi(x,y)$, we easily find that
\bea
w(x,y) &=& c_{10} x+c_{01} y + c_{11} x y +\frac{1}{2} b (x^2-y^2),  \nn\\
K &=& (c_{10}+b x+c_{11} y) p_x-(c_{01}+c_{11} x- b y) p_y +((2 b-c_4)v+c_3 t)p_v + c_4 t p_t, \nn
\eea
with $U(x,y)$ satisfying
$$
(c_{10}+b x+c_{11} y) U_x-(c_{01}+c_{11} x- b y) U_y = 2(b-c_4) U -\frac{1}{2} c_3.
$$
A simple form of solution corresponds to $c_{10}=c_{01}=c_{11}=c_3=0$, giving $U(x,y) = x^{2\left(1-\frac{c_4}{b}\right)} \psi\left(\frac{y}{x}\right)$.  If we choose $c_4=-\frac{1}{2} b$ and $\psi(z) = \alpha z^2+\beta$, we obtain
$$
U(x,y) = \alpha x y^2+\beta x^3, \quad K=\frac{1}{2} b (2 x p_x+2 y p_y+5 v p_v-t p_t).
$$
This is the general H\'enon-Heiles potential (with the harmonic terms suppressed), which is not integrable for arbitrary $\alpha,\, \beta$. However, the choice $\alpha=1,\, \beta=2$ corresponds to an integrable case (associated with the KdV hierarchy \cite{f91-1}), having a second quadratic integral, with Eisenhart Lift
$$
F = p_y (y p_x-x p_y) +y^2 (4 x^2+y^2) p_v^2.
$$
The full Poisson algebra of integrals $p_v, p_t, F, F_K$ is
\bea
&&  \{p_v,p_t\}=\{p_v,F\}=\{p_t,F\}=0,\quad \{p_v,F_K\} = -\frac{5 b}{2}\, p_v^2, \nn\\
&&   \{p_t,F_K\} = \frac{1}{2}b (p_vp_t-2 H),\quad \{F,F_K\} = -b p_v F. \nn
\eea
With $H, p_v, p_t, F$ in involution, the system has retained its Liouville integrability, and has gained one extra integral.

\subsubsection{Starting with $\varphi(x,y) = x^2$}

With this choice of $\varphi(x,y)$, the 2D metric has constant curvature. We easily find that
$$
w(x,y) = c_{01} y +\frac{1}{2}c_{10}(x^2-y^2) +\frac{1}{6} c_{11} (3 x^2 y-y^3) , \qquad b=0.
$$
A simple form of solution corresponds to $c_{01}=c_{11}=c_3=0$, giving $U(x,y) = x^{-\frac{2c_4}{c_{10}}} \psi\left(\frac{y}{x}\right)$.  If we choose $c_{10}=-\frac{1}{2} c_4$ and $\psi(z) = 1+4 z^2$, we obtain
$$
U(x,y) = x^2(x^2+4 y^2), \quad K=-\frac{1}{2} c_4 (x p_x+y p_y+2 v p_v-2t p_t).
$$
This is an integrable potential (in 2 dimensions), having a second quadratic integral, with Eisenhart Lift
$$
F = p_y (x p_x+y p_y) + 16 y (x^2+2y^2) p_v^2.
$$
Since $b=0$, the above $K$ is a first integral.  The full Poisson algebra of integrals $p_v, p_t, K, F$ is
\bea
&&  \{p_v,p_t\}=\{p_v,F\}=\{p_t,F\}=0,\quad \{p_v,K\} = c_4 p_v,   \nn\\
&&  \{p_t,K\} = -c_4 p_t, \quad  \{K,F\} = -\frac{1}{2} c_4 F. \nn
\eea
With $H, p_v, p_t, F$ in involution, the system has retained its Liouville integrability, and has gained one extra integral.

\subsubsection{The Darboux--Koenigs Metric $D_1$ with $\varphi(x,y) = x^{-1}$}

This metric belongs to the classification by Darboux and Koenigs of metrics (see \cite{02-6}) with \underline{only one} Killing vector and two second order Killing tensors (necessarily \underline{not} constant curvature).
We easily find that
$$
w(x,y) = c_{01} y+\frac{1}{3}b (x^2-y^2) .
$$
A simple form of solution corresponds to $c_{01}=c_3=0, c_4=2 b$, giving $U(x,y) = x^{-3} \psi\left(\frac{y}{x}\right)$.  If we choose $\psi(z) = \mu z^{-2}$, we obtain
$$
U(x,y) = \frac{\mu}{x y^2}, \quad K= \frac{2}{3} b (x p_x+y p_y+3t p_t).
$$
This is a {\em superintegrable} potential (in 2 dimensions) (see \cite{02-6}), having two quadratic integrals, with Eisenhart Lifts
$$
F_1 = p_y (y p_x-x p_y) - \frac{8 \mu x}{y^2}p_v^2-\frac{1}{2} y^2 (H+2 p_vp_t),\quad F_2 = p_y^2 +\frac{8 \mu}{y^2} \, p_v^2,
$$
together with
$$
F_3 = \{F_1,F_2\},\quad p_v,\quad p_t,\quad F_K = p_v K+b\, t\, H.
$$
The Poisson relations of these integrals are
\bea
&&  \{p_v,p_t\}=\{p_v,F_i\}=\{p_t,F_i\}=0,\quad \{p_v,F_K\} = 0,\quad \{p_t,F_K\} = -b (H+2 p_v p_t),   \nn\\
&&  \{F_1,F_3\} = -4 (H+2p_vp_t)F_1-6 F_2^2, \quad  \{F_2,F_3\} = 4 (H+2p_vp_t)F_2 , \nn\\
&&  \{F_1,F_K\} = -\frac{2}{3} b p_v F_1,\quad \{F_2,F_K\} = -\frac{4}{3} b p_v F_2,\quad \{F_3,F_K\} = -2 b p_v F_3,\nn
\eea
with the additional relations
$$
\{K,F_1\}=\frac{2}{3}b F_1,\quad \{K,F_2\}=\frac{4}{3} b F_2,\quad \{K,F_3\}= 2 b F_3,\quad \{K,F_K\}=0.
$$
With $H, p_v, p_t, F_1$ in involution, the system has retained its Liouville integrability.  In 4 degrees of freedom, we can have \underline{at most} 7 functionally independent integrals.  We have 7 integrals, but their Jacobian has rank 6 as a result of the algebraic relation
$$
F_3^2 +8 (H+2 p_vp_t)F_1 F_2+4 F_2^3+32 m p_v^2 (H+2 p_vp_t)^2 = 0,
$$
which is a deformation of a similar relation in the $2$--dimensional context.

\section{Energy--Momentum Tensor and Einstein Equations}\label{Energy-Momentum}

We saw in Equation (\ref{dtdtau}) that the Hamiltonian (\ref{GenH}) always has first integrals $p_v$ and $p_t$, corresponding to a pair of commuting Killing vectors
$$
\chi^\mu \partial_\mu=\partial_t, \qquad \xi^\mu \partial_\mu=\partial_v,
$$
where $\partial_\mu=\frac{\partial}{\partial z^\mu}$, the first of which is time--like while the second is null and covariantly constant, which implies (\ref{4dMetric}) belongs to the class of Brinkmann or Pp-wave spacetimes.

The Eisenhart metric (\ref{4dMetric}) does not solve the vacuum Einstein equations unless $U$ and $\log(\varphi)$ are harmonic functions. In this case the original $2D$ metric (\ref{2dMetric}) is actually flat (see (\ref{2dR})).  We saw that the general form of the Einstein tensor of the metric (\ref{4dMetric}) is given by (\ref{4dEinstein}).  With Einstein's equations
\begin{subequations}\label{Einstein-eq}
\bea\label{EE}
R_{\mu\nu}-\frac 12 g_{\mu\nu} R=8\pi T_{\mu\nu},
\eea
the formula (\ref{4dEinstein}) gives the form of $T_{\mu\nu}$ to be
\bea\label{EMT}
T_{\mu\nu}=\frac{1}{2\pi} \Omega \xi_\mu \xi_\nu+\frac{1}{8\pi} \Sigma (\xi_\mu \chi_\nu+\xi_\nu \chi_\mu), \qquad {T^\mu}_\mu=-\frac{1}{8\pi}\Sigma,
\eea
where $\Omega$ and $\Sigma$ are given by
\be\label{OmegaSigma}
\Omega = \varphi \Delta U-R U,\quad \Sigma = R  = \varphi \Delta \log(\varphi),
\ee
with (\ref{EE}) implying that
\be\label{condT}
\nabla^\mu T_{\mu\nu}=0.
\ee
\end{subequations}

Equations (\ref{OmegaSigma}) can be regarded in two different ways. Given the pair $(\Omega,\Sigma)$, (\ref{OmegaSigma}) provide the partial differential equations to fix the metric (\ref{4dMetric}) in a way compatible with the Einstein equations (\ref{EE}). Vice versa, assuming the Lorentzian metric (\ref{4dMetric}) is given, then (\ref{OmegaSigma}) algebraically determine the matter characteristics $(\Omega,\Sigma)$, which fix the energy--momentum tensor (\ref{EMT}).

Let us discuss physical conditions to be imposed on the energy--momentum tensor (\ref{EMT}). The energy density measured by an observer moving along a time--like geodesic $z^\mu=z^\mu(\tau)$ is
\begin{subequations}\label{Energy-cond}
\be\label{energy}
T_{\mu\nu}\frac{dz^\mu}{d\tau} \frac{dz^\nu}{d\tau}=\frac{1}{16\pi}\, t_\tau((v_\tau+4 U t_\tau)\Sigma +2 \Omega t_\tau)= \frac{1}{8\pi}\, t_\tau^2 ({\cal E} \Sigma+\Omega),
\ee
where we have used the form of the matrix (\ref{4dEinstein}), with Equations (\ref{dbydtau}) and (\ref{pt}).  Thus, the weak energy condition $T_{\mu\nu}\frac{dz^\mu}{d\tau} \frac{dz^\nu}{d\tau}\geq 0$ gives
\be\label{weakE}
\Omega+\mathcal{E} \Sigma \geq 0.
\ee
Since for time--like geodesics $\mathcal{E}>0$~\footnote{It follows from $g_{\mu\nu}\frac{dz^\mu}{d\tau} \frac{dz^\nu}{d\tau} =-1$ and Eqs. (\ref{GenH2d}), (\ref{pv}), and (\ref{pt}) that $\mathcal{E}=H^{2d}+\frac{1}{2\kappa^2}>0$. Note that $\mathcal{E}$ can be arbitrarily large.}, it suffices to require
\be\label{wec}
\Omega \geq 0, \qquad \Sigma \geq 0.
\ee
It turns out that (\ref{wec}) also ensures the strong energy condition
$\left( T_{\mu\nu}-\frac 12 g_{\mu\nu} {T^\lambda}_\lambda \right)\frac{dz^\mu}{d\tau} \frac{dz^\nu}{d\tau} \geq 0$, which reduces to
\be
\Omega+\left(\mathcal{E}-\frac{1}{2\kappa^2} \right) \Sigma \geq 0.
\ee
\end{subequations}
The latter is automatically satisfied, as for time--like geodesics $\mathcal{E}>\frac{1}{2\kappa^2}$. Recall that for a geodesic congruence which has a vanishing rotation tensor (hypersurface orthogonal), the strong energy condition implies that the geometry exerts a focussing effect on time--like geodesics. This is a consequence of the Raychaudhuri equation.

We conclude this section with a remark on the physical meaning of $\Omega$ and $\Sigma$. Raising the indices in (\ref{EMT}), one finds two non--vanishing components
\be
T^{tv}=\frac{1}{8\pi} \Sigma, \qquad T^{vv}=\frac{1}{2\pi} \Omega.
\ee
In order to correctly interpret $\Omega$ and $\Sigma$, it proves instructive to make recourse to the flat space
which occurs at $U=0$ and $\varphi=1$
\be\label{dnc}
ds^2=-dt dv+dx^2+dy^2.
\ee
One sees that $t$ and $v$ are actually the double null coordinates. Implementing the coordinate transformation
\be
t=\tilde t+\tilde v, \qquad v=\tilde t-\tilde v,
\ee
one brings (\ref{dnc}) to the standard Minkowski metric, while the energy--momentum tensor acquires the form ($T^{\mu\nu} \partial_\mu \partial_\nu={\tilde T}^{\tilde\mu \tilde\nu} \partial_{\tilde\mu} \partial_{\tilde\nu}$)
\be\label{FEMT}
{\tilde T}^{\tilde t \tilde t}=\frac{1}{8\pi} \left(\Omega+\frac 12 \Sigma \right), \qquad {\tilde T}^{\tilde t \tilde v}=-\frac{1}{8\pi} \Omega,  \qquad {\tilde T}^{\tilde v \tilde v}=\frac{1}{8\pi} \left(\Omega-\frac 12 \Sigma \right).
\ee
At this point $\frac{1}{8\pi} \left(\Omega+\frac 12 \Sigma \right)$ can be identified with the energy density, with  $-\frac{1}{8\pi} \Omega$ being the energy flux density in the direction orthogonal to the $(x,y)$--plane, while $\frac{1}{8\pi} \left(\Omega-\frac 12 \Sigma \right)$ is the only non--vanishing component of the stress tensor.

\section{Hidden Symmetries and Integrable Models}\label{hidden}

We have already seen in Section \ref{first-int} how each first integral of homogeneous degree $m$ corresponds to a Killing tensor of rank $m$ of the metric corresponding to the kinetic energy. This can then be homogenised to sit within the Eisenhart lift, as seen in Section \ref{Hamilton-2d4d}.  In this section we consider some further examples of integrable and superintegrable systems and investigate the physical properties of the corresponding energy-momentum tensors.

\subsection{The Darboux--Koenigs Metrics}

The Darboux--Koenigs metrics, classified in 1898 by Koenigs (see \cite{K}), possess exactly {\em one} Killing vector (hence are {\em not} constant curvature) and a pair of second order Killing tensors (one of which is necessarily functionally dependent).
There are 4 such metrics, characterised by the Killing vector $K=\pa_y$ and the function $\varphi(x,y)$ of (\ref{GenH2d}):
\be\label{DKn}
\begin{array}{ll}
\displaystyle  \varphi_I(x,y) = \frac{1}{x}, & \displaystyle \varphi_{II}(x,y) = \frac{x^2}{1+x^2}, \\[3mm]
\displaystyle \varphi_{III}(x,y) = \frac{e^{2x}}{1+e^x}, & \displaystyle  \varphi_{IV}(x,y) = \frac{\sin^2 x}{a+\cos x},
\end{array}
\ee
where $a>1$.
The existence of the Killing vector and tensors means that the geodesic equations are superintegrable.

Computing the scalar curvature (\ref{OmegaSigma})
\bea
&&
\Sigma_{I}=\frac{1}{x^3}, \qquad  \Sigma_{II}= -\frac{2 (1 + 3 x^2)}{{(1 + x^2)}^3}, \qquad \Sigma_{III}=-\frac{e^{3x}}{{(1+e^x)}^3},
\nonumber\\[2pt]
&&
\Sigma_{IV}=-\frac{2+8a^2+15a\cos{x}+6\cos{2x}+a\cos{3x}}{4{(a+\cos{x})}^3},  \nn
\eea
one concludes that the Eisenhart uplifts of the Darboux--Koenigs metrics of types II, III, and IV
violate the weak energy condition (\ref{wec}),  while type--I models are viable provided the domain $x>0$ is chosen.

In \cite{02-6,KKMW} there is a classification of potential functions $U(x,y)$ for which superintegrability is maintained. In general, $U(x,y)$ breaks the Killing isometry associated with $K=\pa_y$ and makes two second rank Killing tensors functionally independent. Focusing on type--I models, one reveals two options  \cite{02-6}. The first is described by the (separable) Hamiltonian
\be\label{CE}
H=\frac{1}{2x} \left( p_x^2+p_y^2+ b_1 (4 x^2 + y^2)+b_2 + \frac{b_3}{y^2}\right),
\ee
where $b_i\geq 0$ are constants, and two quadratic integrals of motion
$$
F_1=p_y (y p_x-x p_y ) - \frac{y^2 }{2} H +b_1 x y^2-\frac{b_3 x}{y^2}, \qquad  F_2=p_y^2+ b_1 y^2+\frac{b_3}{y^2},
$$
which obey the non--linear algebra jointly with $F_3=\{F_1,F_2 \}$:
\bea
&&
\{F_1,F_3 \}=-4 H F_1-6 F_2^2-4 b_2 F_2+8 b_1 b_3,  \nonumber\\[2pt]
&&
\{F_2,F_3 \}= 4 H F_2 + 16 b_1 F_1 .  \nn
\eea
Demanding $\Omega$ to be non--negative
$$
\Omega=\frac{b_1  (-2 x^2 + y^2)y^4+ b_2 y^4  +b_3 (6 x^2 + y^2) }{2 x^4 y^4}\geq 0,
$$
one is led to set $b_1=0$. The region $x>0$ is invariant under the Hamiltonian flow:  supposing $x>0$ at $t=0$, then $H=E>0$ at $t=0$ and therefore, for all $t\geq 0$.  Hence, noting that all terms within the parentheses of (\ref{CE}) are positive definite, we have $x>0$ for all $t\geq 0$.

Another type--I model, which is compatible with the weak energy conditions (\ref{wec}), is governed by the (separable) Hamiltonian
\be\label{dk1b}
H=\frac{1}{2x} \left( p_x^2+p_y^2+ b_1+b_2 (x^2+y^2)\right),
\ee
where $b_1,b_2\geq0$ are constants, and integrals of motion
$$
F_1=p_x p_y-y H+b_2 x y, \qquad  F_2=p_y^2 +b_2 y^2.
$$
Along with $F_3=\{F_1,F_2 \}$, they form the Poisson algebra
$$
\{F_1,F_3 \}=2H^2-4 b_2 F_2-2 b_1 b_2, \qquad   \{F_2,F_3 \}=4 b_2 F_1.
$$
The corresponding $\Omega$ is positive--definite
$$
\Omega=\frac{b_1 +  b_2 (x^2 + y^2)}{2 x^4}.
$$
Using the same argument as before, the domain $x>0$ is invariant under the Hamiltonian flow, so $x>0$ for all $t\geq 0$.

We will have more to say about separable systems in Section \ref{simple-sep}.

\subsection{A Superintegrable System on the Two--Dimensional Sphere}

Consider the $2$--dimensional superintegrable systems on a constant curvature space, admitting two quadratic constants of the motion. To have positive $\Sigma$  we focus on models on the two--dimensional sphere (classified in \cite{KKPM}), which are invariant under the rotation group, whose infinitesimal generators form a simple Poisson algebra:
\be\label{rot-alg}
L_1 = -p_\phi \cos \phi \cot \theta - p_\theta \sin \phi,\quad L_2 = p_\theta \cos \phi - p_\phi \cot \theta \sin \phi,\quad L_3 = p_\phi,
\ee
satisfying $\{L_i,L_j\}=\epsilon_{ijk}L_k$, with Casimir $L_1^2+L_2^2+L_3^3=p_\theta^2+\sin^{-2}\theta p_{\phi}^2$.

As an example, consider the system $(S9)$ in \cite{KKPM}, written in spherical coordinates
\begin{subequations}
\bea
H &=&\frac 12 \left(p_\theta^2+\sin^{-2}\theta p_{\phi}^2 \right) +U(\theta,\phi), \nn\\[-2mm]
&&  \label{sp}  \\[-2mm]
U(\theta,\phi)&=&
\frac 12 \left(\frac{1}{\sin^2 \theta} \left(\frac{\alpha^2}{\sin^2 \phi}+\frac{\beta^2}{\cos^2 \phi} \right)+\frac{\gamma^2}{\cos^2 \theta} \right),  \nn
\eea
where $(\theta,p_\theta)$ and $(\phi,p_\phi)$ are canonical pairs and $\alpha$, $\beta$, $\gamma$ are free parameters (coupling constants). The model is characterized by two quadratic integrals of motion \cite{KKPM,GNS}
\bea
&&  F_1 = L_3^2+f_1 = p_\phi^2+\frac{\alpha^2}{\sin^2 \phi}+\frac{\beta^2}{\cos^2 \phi},  \nn\\[-2mm]
&&    \label{com1}     \\[-2mm]
&&  F_2 = L_1^2+f_2= {(p_\phi \cos{\phi} \cot{\theta}+p_\theta \sin{\phi})}^2+{\left(\alpha \frac{\cot{\theta}}{\sin{\phi}} +\gamma \sin{\phi} \tan{\theta}\right)}^2,  \nn
\eea
which along with $H$ form a functionally independent set.  We define $F_3=\frac{1}{4} \{F_1,F_2\}= L_1L_2L_3+\mbox{``first order terms''}$, where the leading order term is determined by the Poisson relations of $L_i$.  This is not functionally independent of $H, F_1, F_2$ and satisfies the polynomial constraint (\ref{F32}) below
\bea
F_3^2 &=& {\cal P}(F_1,F_2,H) =  F_1 F_2 (2 H-F_1-F_2) -(\a-\g)^2 F_1^2-(\a^2-\b^2-4 \a \g + \g^2) F_1 F_2 \nn\\
  && +4 \a (\a-\g) F_1 H +2 (\a^2-\b^2)F_2 H -4 \a^2 H^2 +2 \g (\a-\g) (\a^2-\b^2-\a \g) F_1\nn \\
  && \qquad  -(\a^2-\b^2)\g^2 F_2-4 \a \g (\a^2-\b^2-\a \g)H-\g^2(\a^2-\b^2-\a\g)^2.  \label{F32}
\eea
The leading order term is easily found by considering $L_1^2L_2^2L_3^2$.  We can use this constraint to determine the final two relations of our Poisson algebra:
\bea
\{F_1,F_3\} &=& 2 \frac{\pa {\cal P}}{\pa F_2} = 2(4\a \g F_1-(\a^2-\b^2+F_1)(\g^2+F_1-2 H) -2 F_1F_2),\label{F1F3}  \\
\{F_2,F_3\} &=& -2 \frac{\pa {\cal P}}{\pa F_1} = -2((2H-2F_1-F_2-(\a^2-\b^2-4\a \g+\g^2))F_2-2(\a-\g)^2 F_1 \nn\\
  &&  \hspace{3cm} +4 \a (\a-\g) H +2 \g(\a-\g)(\a^2-\b^2-\a\g)),\label{F2F3}
\eea
\end{subequations}

Whilst (\ref{sp}) is not explicitly written in the conformally flat form of (\ref{GenH2d}), we can clearly replace the original kinetic energy by that on a two--dimensional sphere, with metric $d\theta^2+\sin^2{\theta} d\phi^2$, replace $U(x,y)$ by $U(\theta,\phi)$, and impose the Einstein equations to find $\Sigma=2$ and
\bea
\Omega &=& \frac{\alpha^2 (5 + \cos{2\phi})}{2\sin^4{\theta}\sin^4{\phi}}+\frac{\beta^2 (5-\cos{2\phi})}{2 \sin^4{\theta} \cos^4{\phi}}+\frac{\gamma^2 (2-\cos{2\theta})}{\cos^4{\theta}}
\nonumber\\[2pt]
&&
\hspace{4cm}  +\frac{ 2(\alpha^2 + \beta^2 + (\alpha^2 - \beta^2) \cos{2\phi}) \cos{2\theta}}{\sin^2{2\phi}\sin^4{\theta}}.  \nn
\eea
Combining the terms involving $\alpha^2$, $\beta^2$ and $\gamma^2$ separately, one can verify that $\Omega$
is positive definite. Killing tensors associated with constants of the motion (\ref{com1}) arise in the usual way.

\subsection{Some Liouville Integrable Systems in Cartesian Coordinates}\label{hid-Liouv}

Superintegrable systems have very rigid choices of both $\varphi(x,y)$ and $U(x,y)$, so leave no room for manoeuvering $\Sigma$ and $\Omega$ to be non-negative.  On the other hand, Liouville integrable (including {\em separable}) systems can have potentials with arbitrary functions, which can be judiciously chosen as in the examples below.

\subsubsection{A Simple Separable System}\label{simple-sep}

Consider the Poisson commuting pair
\be\label{H1}
H=\frac{1}{2} \varphi(x) (p_x^2+p_y^2+u(y)),\quad F = p_y^2+u(y),
\ee
where $\varphi(x)$ and $u(y)$ are positive definite functions. The Eisenhart metric (\ref{4dMetric}) associated with (\ref{H1}) admits the second rank Killing tensor corresponding to $F$ (of (\ref{F4d})), whose upper index form has non-zero components $f^{yy}=1, f^{vv}=4 u(y)$, which can be lowered with (\ref{4dMetric}) to give non--vanishing components
$$
f_{tt}=u(y), \qquad f_{yy}={\varphi(x)}^{-2}.
$$
If we choose $\varphi(x)=x^{-2}$ and $u(y)=y^2$, then from (\ref{OmegaSigma}) we find
$$
\Omega =\frac{x^2+2y^2}{x^6}, \qquad \Sigma =\frac{2}{x^4},
$$
which clearly obey the weak energy conditions (\ref{wec}).

With these choices of $\varphi(x)$ and $U(x,y)=\frac{1}{2}\varphi(x)u(y)$, equations (\ref{w-eq}) give the conformal symmetry
$$
K = \frac{1}{2}\, b\, (x p_x+y p_y+2 v p_v+2 t p_t)
$$
and the additional quadratic integral $F_K$ (see (\ref{FK})), also defining a second rank Killing tensor, together with the following (non-zero) Poisson relations:
\bea
&&  \{K,H\}=2 b H,\;\;\; \{K,F\}= b F,\;\;\; \{K,F_K\}= b F_K,\;\;\; \{K,p_v\}= b p_v,\;\;\; \{K,p_t\}= b p_t,  \nn\\[3mm]
&& \{F,F_K\} = -b p_v F ,\;\;\; \{F_K,p_v\}= b p_v^2,\;\;\; \{F_K,p_t\} = b(H+p_vp_t).  \nn
\eea

\subsubsection{The Flat Metric with Quartic Potential}

Consider the Poisson commuting pair
\be\label{H-16121}
H=\frac{1}{2} (p_x^2+p_y^2)+ 16 x^4 +12 x^2 y^2 +y^4,\quad F = p_y(yp_x-xp_y) + 4 xy^2(2 x^2+y^2),
\ee
which is a well known integrable system, separable in parabolic coordinates.  In fact, there are infinitely many homogeneous polynomial potentials, including the H\'enon--Heiles potential of Section \ref{HH-pot}, which are separable in parabolic coordinates.  The even degree polynomials give an energy-momentum tensor (of the Eisenhart lift) satisfying the weak energy condition.  The above quartic potential gives (using (\ref{OmegaSigma})):
$$
\Omega = 36(6x^2+y^2), \qquad \Sigma = 0,
$$
which clearly obey the weak energy conditions (\ref{wec}).

With these choices of $\varphi(x)$ and $U(x,y)$, equations (\ref{w-eq}) give the conformal symmetry
$$
K =  b\, (x p_x+y p_y+ 3 v p_v - t p_t)
$$
and the additional quadratic integral $F_K$ (see (\ref{FK})), with both $F$ and $F_K$ defining second rank Killing tensors.  These functions satisfy the following (non-zero) Poisson relations:
\bea
&&  \{K,H\}=2 b H,\;\;\; \{K,F\}= b F,\;\;\; \{K,F_K\}= 3b F_K,\;\;\; \{K,p_v\}= 3b p_v,\;\;\; \{K,p_t\}= -b p_t,  \nn\\[3mm]
&& \{F,F_K\} = -b p_v F ,\;\;\; \{F_K,p_v\}= 3b p_v^2,\;\;\; \{F_K,p_t\} = b(H-p_vp_t).  \nn
\eea

\subsubsection{Specific Case from Section \ref{given-w}}

In Section \ref{given-w}, we specified $w(x,y)=x^2-y^2$ (in the definition of the conformal symmetry $K$) and derived the corresponding general form of $\varphi(x,y)$ and $U(x,y)$ in terms of 2 arbitrary functions of a single variable: $A\left(\frac{y}{x}\right)$ and $B\left(\frac{y}{x}\right)$.  Equations (\ref{OmegaSigma}) give $\Omega$ and $\Sigma$ as differential expressions in $A$ and $B$.  We don't propose here to analyse these expressions to determine all specific forms of $A$ and $B$ for which the weak energy conditions (\ref{wec}) is satisfied, but clearly there would be many viable cases.  If we choose $A\left(\frac{y}{x}\right)\equiv 1,\, b=2$, then we have the flat metric with $\varphi(x,y)=1$, so $\Sigma=R=0$.  If we now specify the two parameters $c_3=0,c_4=-2$, then
$$
\Omega = (x^2+y^2) B''\left(\frac{y}{x}\right)-6 x y B'\left(\frac{y}{x}\right)+12 x^2 B\left(\frac{y}{x}\right),
$$
so just need to choose the form of $B\left(\frac{y}{x}\right)$ so that this expression is non--negative for all $(x,y)$.  For example
$$
B(z) = d_1+d_2 z^2+d_3 z^4 \quad\Rightarrow\quad \left\{ \begin{array}{l}
                                                           U(x,y)=d_1 x^4+d_2 x^2 y^2+d_3 y^4,  \\[2mm]
                                                           \Omega = 2 (6d_1+d_2)x^2+2 (d_2+6d_3) y^2,
                                                           \end{array}   \right.
$$
so $\Omega$ is positive definite whenever the two coefficients are positive.  In all cases, this has conformal symmetry
$$
K =  2 (x p_x+y p_y+ 3 v p_v - t p_t)\quad\mbox{with}\quad \{K,H\} = 4 H,
$$
regardless of whether or not we choose an integrable case, such as $(d_1,d_2,d_3)=(16,12,1)$, as above.

\section{Conclusion}\label{conclude}

In this paper we have considered the Eisenhart lift of a fairly general $2$--dimensional metric, with particular interest in comparing the Hamiltonian properties of the $2D$ system and its $4D$ lift.
Whilst the original $2D$ metric (\ref{2dMetric}) was conformally flat, its Lorentzian counterpart (\ref{4dMetric}) fails to be so. Computing the Weyl tensor, one can verify that its vanishing requires $\Delta \log (\varphi)=0$, which in turn implies that the metric (\ref{2dMetric}) is flat. This makes the cosmological applications in the spirit of a recent work \cite{CGGH} problematic.  Furthermore, it is worth recalling that reductions of the Goryachev--Chaplygin and Kovalevskaya tops, which are obtained by discarding a cyclic variable, result in $2D$ integrable systems in curved space possessing cubic and quartic integrals of motion, respectively. The corresponding Eisenhart metrics and higher rank Killing tensors were constructed in \cite{GHKW}. One can verify that, while $\Sigma>0$, $\Omega$ fails to be positive definite in the whole domain thus ruling out these examples from the physically acceptable list.

In our derivation of the conformal symmetry (\ref{Ksol}), we made several restrictions on the coefficients $a_i(x,y,v,t)$, so an obvious question is whether a more general solution can be found, or even some algebra of independent conformal symmetries can be found.  Even with these restrictions, we found that a {\em given} function $w(x,y)$ led to $\varphi(x,y)$ and $U(x,y)$, depending upon arbitrary functions.  Particular choices can lead to energy--momentum tensors obeying the weak energy condition.  There is therefore the question of how to choose these functions in a more systematic way and, if possible, to classify these.

It would be interesting to generalise the analysis in this work to $d>2$ mechanics on conformally flat backgrounds. The first problem is that the Einstein tensor would not have such a simple decomposition as (\ref{4dEinstein}), so the energy--momentum tensor would be more difficult to analyse.  Secondly, the calculation of the conformal symmetry, as in Section \ref{conform-sym}, would be considerably more complex.

Another open problem is whether Lorentzian metrics admitting third or higher rank Killing tensors, linked to the work in \cite{VDS,V} give rise to physically meaningful solutions.

\subsubsection*{Acknowledgements}

This work was supported by the Tomsk Polytechnic University competitiveness enhancement program.


\end{document}